\documentclass[12pt,english]{article}
\usepackage[T1]{fontenc}
\usepackage{a4wide}
\usepackage{amsmath}
\usepackage{babel}
\usepackage{graphics}
\usepackage{amssymb}
\usepackage{subfigure}

\makeatletter

\providecommand{\LyX}{L\kern-.1667em\lower.25em\hbox{Y}\kern-.125emX\@}

\newcommand{\eqalign}[1]{
\null \,\vcenter {\openup \jot \ialign {\strut \hfil $\displaystyle {
##}$&$\displaystyle {{}##}$\hfil \crcr #1\crcr }}\,}
\usepackage{subfigure}
\newcommand{\be}{\begin{equation}}
\newcommand{\ee}{\end{equation}}

\@addtoreset{equation}{section}

\@addtoreset{figure}{section}

\@addtoreset{table}{section}                                                    

\makeatother
\begin{document}

\title{{\normalsize \begin{flushright}\normalsize{ITP--Budapest Report 571}\end{flushright}\vspace{1cm}}
The spectrum of boundary states in sine-Gordon model with integrable
boundary conditions}

\author{Z. Bajnok, L. Palla, G. Takács and G.Zs. Tóth}

\maketitle
{\centering \emph{Institute for Theoretical Physics }\\
\emph{Roland Eötvös University, }\\
\emph{H-1117 Budapest, Pázmány sétány 1/A, Hungary}\par}

\begin{abstract}
The bound state spectrum and the associated reflection factors are
determined for the sine-Gordon model with arbitrary integrable boundary
condition by closing the bootstrap. Comparing the symmetries of the
bound state spectrum with that of the Lagrangian it is shown how one
can {}``derive{}'' the relationship between the UV and IR parameters
conjectured earlier. 
\end{abstract}
{\par\centering PACS codes: 64.60.Fr, 11.10.Kk   
\par}
\newpage
\section{Introduction}

The sine-Gordon model is one of the most extensively studied quantum
field theories. The interest stems partly from the wide range of
applications that extend from particle physics to condensed matter
systems and partly from the fact that many of the interesting physical
quantities can be computed exactly due to its integrability. 
As was argued in \cite{GZ,Skly}, the boundary version of sine-Gordon model:
\begin{equation}
S=\int _{-\infty }^{\infty }dt\int _{-\infty }^{0}dx{\cal {L}}_{SG}-\int _{-\infty }^{\infty }dtV_{B}(\Phi _{B}),\qquad \quad {\cal {L}}_{SG}=\frac{1}{2}(\partial _{\mu }\Phi )^{2}-\frac{m^{2}}{\beta ^{2}}(1-\cos (\beta \Phi )),
\end{equation}
 (where \( \Phi (x,t) \) is a scalar field, \( \beta  \) is a real
dimensionless coupling and \( \Phi _{B}(t)=\Phi (x,t)|_{x=0} \))
preserves the integrability of the bulk theory if the boundary potential
is chosen as \[
V_{B}(\Phi _{B})=M_{0}\left( 1-\cos \left( \frac{\beta }{2}(\Phi _{B}-\phi _{0})\right) \right) ,\]
 where \( M_{0} \) and \( \phi _{0} \) are free parameters. As a
result of the boundary potential the scalar field satisfies the boundary
condition: \begin{equation}
\label{hatfel}
\partial _{x}\Phi |_{x=0}=-M_{0}\frac{\beta }{2}\sin \left( \frac{\beta }{2}(\Phi _{B}-\phi _{0})\right) .
\end{equation}

A novel feature of the boundary sine-Gordon model (BSG) is the
appearance of a complicated spectrum of boundary bound states (BBS) in
addition to the well known bulk ones \cite{GZ}-\cite{patr},
\cite{skod}.  The complete spectrum of these bound states and a full
explanation of all the poles in the reflection factors are known in
two special cases only: in the case of Dirichlet boundary conditions
(which corresponds to taking \( M_{0}\rightarrow \infty \), \( \Phi
_{B}(t)\equiv \phi _{0} \)) they are given in \cite{patr}, while for
Neumann boundary condition (when \( M_{0}=0 \), thus \( \phi _{0} \)
becoming irrelevant) they are presented in \cite{saj}. In the general
case the reflection factors and the bound state spectrum depend on two
\lq infrared parameters' \( \eta \) and \( \vartheta \), that are
determined somehow by the \lq ultraviolet parameters' \( M_{0} \) and
\( \phi _{0} \) appearing in the Lagrangian. However, the precise
relationship between them (we call it UV -- IR relation) is known only
for Dirichlet boundary conditions \cite{GZ}. In contrast in the
boundary sinh-Gordon model the UV -- IR relation was determined by
comparing the WKB and bootstrap spectra \cite{Ed}.\footnote{%
It should be remarked, that Al. B. Zamolodchikov has also calculated (but
not yet published) this relation -- in both the sinh-Gordon and
sine-Gordon cases -- using quite different arguments \cite{Zamm}.  }

The purpose of this paper is twofold: first, by using the bootstrap
principle, we determine the bound state spectrum, the associated
reflection factors and explain their poles in the general case.  We
wish to point out that much of the structure of the bootstrap in the
Dirichlet limit \cite{patr} carries over to the general case with some
modifications, however; some care must be taken for special values of
the parameters (of which the case of the Neumann boundary condition is
an example). The inductive analysis is based on the assumption, that
any pole of the reflection factors in the physical strip that is
not explained by  Coleman-Thun diagrams \cite{CT} corresponds to bound
states\footnote{The
importance of these 
diagrams in the context of boundary bootstrap was first emphasized in
\cite{bCT}.}.
While we also tried to achieve a more systematic and
complete investigation of Coleman-Thun diagrams
 than the one presented
in \cite{patr}, we relied very much on their results in this respect
as well.

Second, by comparing the parameter dependencies of some patterns (such as
global symmetries and ground state sequences) in the bootstrap
solution and in the classical theory we present arguments for the UV
-- IR relation in sine-Gordon model. Naturally, this relation is the
same as the one obtained by appropriate analytic continuation from the
sinh-Gordon case.

The outline of this paper is the following: First we describe the
relevant properties of the classical theory. Then we collect the ground
state reflection factors obtained in \cite{GZ, gosh}. Analysing the
pole structure of the reflection factors we determine the fundamental
domain of the parameters. In the next step we explain the pole structure
of both the solitonic and the breather reflection factors on the ground
state and on the first excited boundaries. These results generalize
straightforwardly giving the full spectrum of boundary bound states
and reflection factors. Finally we comment on some special cases and
give the arguments for the UV-IR relation.

\section{Classical considerations}

Let us focus on the classical theory first. As a consequence of the
boundary potential the discrete symmetries of the bulk theory are
broken in general:  

\begin{itemize}
\item the \( \Phi \mapsto \frac{2\pi }{\beta }-\Phi  \) transformation
(that changes solitons to antisolitons and vice versa) now maps
either \( \phi _{0}\mapsto \frac{2\pi }{\beta }-\phi _{0} \) or \( M_{0}\mapsto -M_{0} \),
thus making the related boundary theories equivalent. Clearly the
symmetry is restored either for \( \phi _{0}=\frac{\pi }{\beta } \)
or for \( M_{0}=0 \).
\item The other \( \mathbb Z_{2} \) symmetry which maps \( \Phi  \) to
\( -\Phi  \) induces the \( \phi _{0}\mapsto -\phi _{0} \) transformation
on the boundary parameters and is realized in the \( \phi _{0}=0 \)
case.  
\end{itemize}
Collecting all the possible equivalences between the boundary parameters
their fundamental domain turns out to be:
\[
0\leq M_{0}\leq \infty \quad ;\qquad 0\leq \phi _{0}\leq \frac{\pi }{\beta }\]
In the general case the classical ground state has to satisfy \[
\lim _{x\mapsto -\infty }\Phi (x,t)=\left\{ \begin{array}{c}
0\\
\frac{2\pi }{\beta }
\end{array}\right. \, \, ,\]
 in addition to (\ref{hatfel}). More precisely there are two possibilities
corresponding to these two choices and the one with the lower energy
is the ground state. It is important to notice that the upper/lower
choice can be realized by a static bulk soliton/antisoliton \lq standing
at the right place': i.e. by choosing \( \Phi \equiv \Phi _{s}(x,a) \)
or \( \Phi \equiv \Phi _{\bar{s}}(x,\bar{a}) \) for \( x\leq 0 \),
where \[
\Phi _{s}(x,a)=\frac{4}{\beta }{\textrm{arctg}}(e^{m(x-a)}),\qquad \qquad \Phi _{\bar{s}}(x,\bar{a})=\frac{2\pi }{\beta }-\Phi _{s}(x,\bar{a}),\]
 and \( a \) (\( \bar{a} \)) is obtained from eq.(\ref{hatfel}). (This fits into the scheme of paper \cite{SSS}
where the authors described  the classical scattering of solitons by including not only a mirror image of 
the soliton but also a standing soliton background).
Indeed this way we find the location of the soliton/antisoliton \[
\sinh (ma)=\frac{\frac{4m}{M_{0}\beta ^{2}}+\cos (\frac{\beta }{2}\phi _{0})}{\sin (\frac{\beta }{2}\phi _{0})},\quad \qquad \sinh (m\bar{a})=\frac{\frac{4m}{M_{0}\beta ^{2}}-\cos (\frac{\beta }{2}\phi _{0})}{\sin (\frac{\beta }{2}\phi _{0})}.\]
 (They are obtained from each other by \( \phi _{0}\leftrightarrow \frac{2\pi }{\beta }-\phi _{0} \)).
The energies of these two solutions can be written as \begin{eqnarray}
E_{s}(M_{0},\phi _{0}) & \equiv  & E_{\textrm{bulk}}+V_{B}=\frac{4m}{\beta ^{2}}+M_{0}-M_{0}R(+)\nonumber \\
E_{\bar{s}}(M_{0},\phi _{0}) & = & \frac{4m}{\beta ^{2}}+M_{0}-M_{0}R(-)=E_{s}(M_{0},\frac{2\pi }{\beta }-\phi _{0})\label{energiak} 
\end{eqnarray}
 where we introduced \( R(\pm )=\left[ 1\pm \frac{8m}{M_{0}\beta ^{2}}\cos (\frac{\beta }{2}\phi _{0})+\frac{16m^{2}}{M_{0}^{2}\beta ^{4}}\right] ^{1/2} \).
From the difference \[
E_{\bar{s}}(M_{0},\phi _{0})-E_{s}(M_{0},\phi _{0})=\frac{16m}{\beta ^{2}}\frac{\cos (\frac{\beta }{2}\phi _{0})}{(R(+)+R(-))}\]
 we see that for \( 0\leq \phi _{0}<\frac{\pi }{\beta } \) the soliton
generates the ground state and the antisoliton the first excited one,
at \( \phi _{0}=\frac{\pi }{\beta } \) they become degenerate and
at \( \phi _{0}>\frac{\pi }{\beta } \) they swap. From eq.(\ref{energiak})
we find in the \( \mathbb Z_{2} \) symmetric limit with \( \phi _{0}\to 0+ \)
\footnote{This limit is not smooth. If we set $\phi_{0}=0$ from the start, 
then the two solutions (\(\Phi_1\equiv\frac{2\pi}{\beta}\) and \(\Phi_2\equiv 0\)) 
of the case $M_{0}<\frac{4m}{\beta ^{2}}$ extend to the case $M_{0}>\frac{4m}{\beta ^{2}}$.
In the latter case however, the one higher in energy, $\Phi_1$, can "decay" into $\Phi_2$   
by emitting a moving single soliton. (We thank the referee for pointing out this). This is consistent with
the quantum theory where no excited boundary state exists neither 
in this domain of the parameters nor in the 
corresponding $\phi_0\to 0+$ limiting case.}   
\begin{equation}
\label{mcritcl}
E_{s}=0,\qquad \qquad E_{\bar{s}}=\left\{ \begin{array}{c}
2M_{0}\qquad M_{0}<\frac{4m}{\beta ^{2}}\\
\frac{8m}{\beta ^{2}}\qquad M_{0}>\frac{4m}{\beta ^{2}}
\end{array}\right. \, \, ,
\end{equation}
 while for \( \phi _{0}=\frac{\pi }{\beta } \) the degenerate energies
can be written \[
E_{s}=E_{\bar{s}}=\frac{4m}{\beta ^{2}}+M_{0}-M_{0}\sqrt{1+\frac{16m^{2}}{M_{0}^{2}\beta ^{4}}}\quad \longrightarrow \quad \left\{ \begin{array}{c}
0\qquad M_{0}\rightarrow 0\\
\frac{4m}{\beta ^{2}}\qquad M_{0}\rightarrow \infty 
\end{array}\right. \, \, .\]

\section{Boundary spectrum from bootstrap principle}

\subsection{Bulk scattering properties}

\begin{figure}
\subfigure[Breather fusion]{\resizebox*{!}{5cm}{\includegraphics{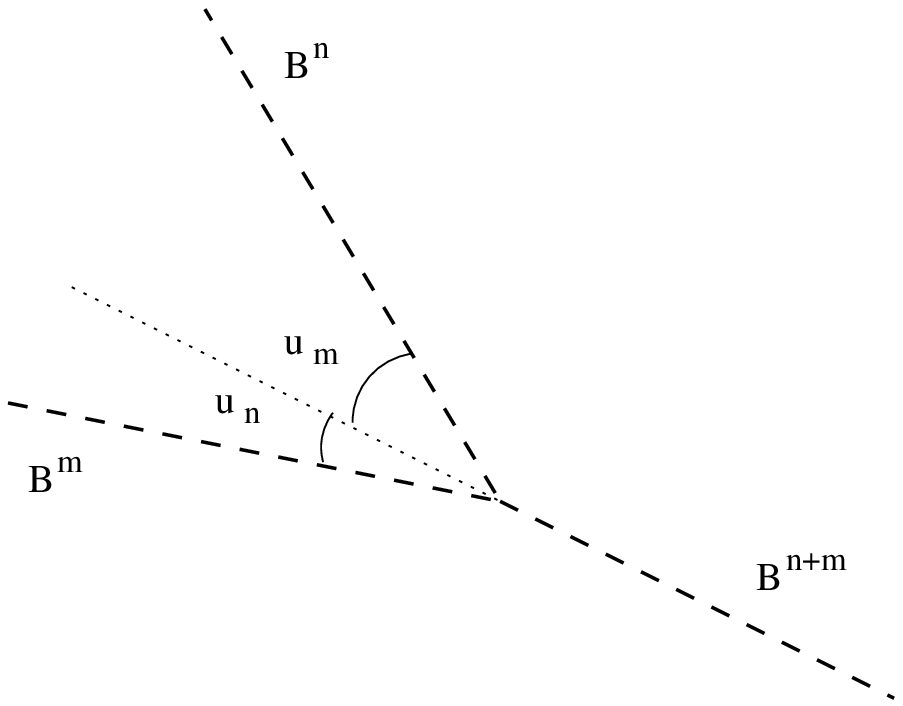}}} 
~~~~~~~~~~\subfigure[Soliton and antisoliton fuse to a breather]{\resizebox*{!}{5cm}{\includegraphics{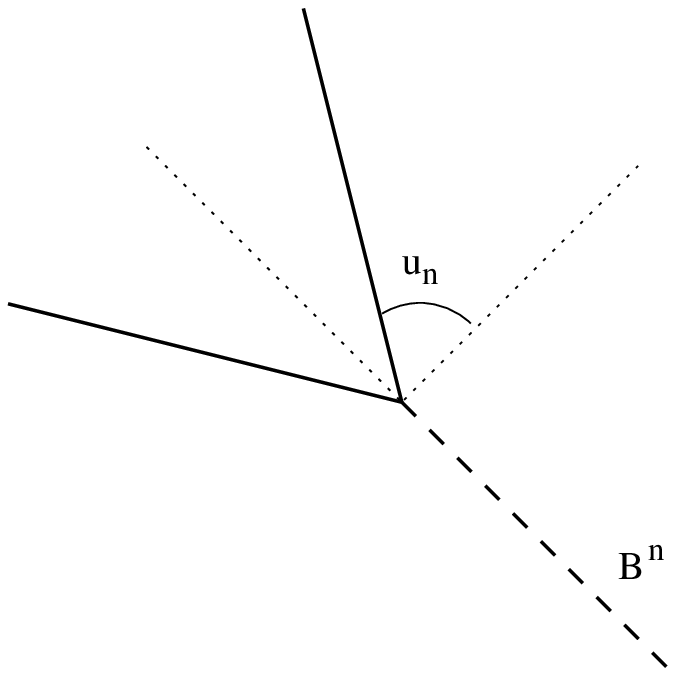}}} 
\caption{\label{bulk_fusion} Bulk fusion processes}
\end{figure}

In the bulk sine-Gordon model any scattering amplitude factorizes
into a product of two particle scattering amplitudes, from which the
independent ones in the purely solitonic sector are \cite{ZZ} 
\begin{eqnarray}
a(u)= & S^{++}_{++}(u)=S_{--}^{--}(u)= & -\prod ^{\infty }_{l=1}\left[ \frac{\Gamma (2(l-1)\lambda -\frac{\lambda u}{\pi })\Gamma (2l\lambda +1-\frac{\lambda u}{\pi })}{\Gamma ((2l-1)\lambda -\frac{\lambda u}{\pi })\Gamma ((2l-1)\lambda +1-\frac{\lambda u}{\pi })}/(u\to -u)\right] \nonumber \\
b(u)= & S^{+-}_{+-}(u)=S_{-+}^{-+}(u)= & \frac{\sin (\lambda u)}{\sin (\lambda (\pi -u))}a(u)\quad \qquad ;\qquad \lambda =\frac{8\pi }{\beta ^{2}}-1\nonumber \\
c(u)= & S^{-+}_{+-}(u)=S_{-+}^{+-}(u)= & \frac{\sin (\lambda \pi )}{\sin (\lambda (\pi -u))}a(u)\quad \qquad ;\qquad u=-i\theta \, \, \, ,\label{abc} 
\end{eqnarray}
Since we are concentrating on the bound state poles located at purely
imaginary rapidities we use the variable \( u \) instead of \( \theta  \)
and refer to it as the rapidity from now on. The other scattering
amplitudes can be described in terms of the functions\[
\{y\}=\frac{\left( \frac{y+1}{2\lambda }\right) \left( \frac{y-1}{2\lambda }\right) }{\left( \frac{y+1}{2\lambda }-1\right) \left( \frac{y-1}{2\lambda }+1\right) }\quad ,\quad (x)=\frac{\sin \left( \frac{u}{2}+\frac{x\pi }{2}\right) }{\sin \left( \frac{u}{2}-\frac{x\pi }{2}\right) }\, \, \, ,\, \, \, \{y\}\{-y\}=1\quad ,\, \, \{y+2\lambda \}=\{-y\}\]
as follows. For the scattering of the breathers \( B^{n} \) and \( B^{m} \)
with \( n\geq m \) and relative rapidity \( u \) we have \cite{ZZ}

\[
S^{n\, m}(u)=S^{n\, m}_{n\, m}(u)=\{n+m-1\}\{n+m-3\}\dots \{n-m+3\}\{n-m+1\}\, \, \, ,\]
while for the scattering of the soliton (antisoliton) and \( B^{n} \)
we have \cite{ZZ}

\[
S^{n}(u)=S_{+\, n}^{+\, n}(u)=S_{-\, n}^{-\, n}(u)=\{n-1+\lambda \}\{n-3+\lambda \}\dots \left\{ \begin{array}{c}
\{1+\lambda \}\quad \textrm{if }n\textrm{ is even}\\
-\sqrt{\{\lambda \}}
\quad \textrm{if }n\textrm{ is odd}\, \, \, .
\end{array}\right. \]
\footnote{Observe that $\{\lambda \}$ is a complete square}
All the poles in the physical strip of the scattering amplitudes originate
from virtual processes either in the forward or in the cross channel
of diagrams \ref{bulk_fusion} (a,b), where the useful definition \[
u_{n}=\frac{n\pi }{2\lambda }\]
 is also introduced. Time develops from top to bottom and solitons
or antisolitons are denoted by solid lines while breather by dashed
ones. To each such process a coupling as \( f_{n\, m}^{n+m} \)or
\( f_{+-}^{n} \) can be attributed and it is known that \( f_{+-}^{n}=(-1)^{n}f_{-+}^{n} \). 

\subsection{Ground state reflection factors}

The most general reflection factor - modulo CDD-type factors - of
the soliton antisoliton multiplet \( |s,\bar{s}\rangle  \) on the
ground state boundary, denoted by \( |\, \rangle  \), satisfying
the boundary versions of the Yang Baxter, unitarity and crossing equations
was found by Ghoshal and Zamolodchikov \cite{GZ}: \begin{eqnarray}
R(\eta ,\vartheta ,u)=\left( \begin{array}{cc}
P^{+}(\eta ,\vartheta ,u) & Q(\eta ,\vartheta ,u)\\
Q(\eta ,\vartheta ,u) & P^{-}(\eta ,\vartheta ,u)
\end{array}\right) =\, \ \ \ \ \ \ \ \ \ \ \ \ \ & & \nonumber \\
\left( \begin{array}{cc}
P_{0}^{+}(\eta ,\vartheta ,u) & Q_{0}(u)\\
Q_{0}(u) & P_{0}^{-}(\eta ,\vartheta ,u)
\end{array}\right) R_{0}(u)\frac{\sigma (\eta ,u)}{\cos (\eta
)}\frac{\sigma (i\vartheta ,u)}{\cosh (\vartheta )}\, \, \, & & ;\nonumber \\
P_{0}^{\pm }(\eta ,\vartheta ,u)=\cos (\lambda u)\cos (\eta )\cosh
(\vartheta )\mp \sin (\lambda u)\sin (\eta )\sinh (\vartheta );
 & &  \nonumber \\
Q_{0}(u)=-\sin (\lambda u)\cos (\lambda u) , \ \ \ \ \ \ \ \ \ \ \ \  \ \ \ \  & \label{Rsas} 
\end{eqnarray}
where \( \eta  \) and \( \vartheta  \) are the two real parameters
characterizing the solution, \[
R_{0}(u)=\prod ^{\infty }_{l=1}\left[ \frac{\Gamma (4l\lambda -\frac{2\lambda u}{\pi })\Gamma (4\lambda (l-1)+1-\frac{2\lambda u}{\pi })}{\Gamma ((4l-3)\lambda -\frac{2\lambda u}{\pi })\Gamma ((4l-1)\lambda +1-\frac{2\lambda u}{\pi })}/(u\to -u)\right] \]
 is the boundary condition independent part and \[
\sigma (x,u)=\frac{\cos x}{\cos (x+\lambda u)}\prod ^{\infty }_{l=1}\left[ \frac{\Gamma (\frac{1}{2}+\frac{x}{\pi }+(2l-1)\lambda -\frac{\lambda u}{\pi })\Gamma (\frac{1}{2}-\frac{x}{\pi }+(2l-1)\lambda -\frac{\lambda u}{\pi })}{\Gamma (\frac{1}{2}-\frac{x}{\pi }+(2l-2)\lambda -\frac{\lambda u}{\pi })\Gamma (\frac{1}{2}+\frac{x}{\pi }+2l\lambda -\frac{\lambda u}{\pi })}/(u\to -u)\right] \]
describes the boundary condition dependence. Note that the topological
charge may be changed by two in these reflections, thus the parity
of the soliton number is conserved. 

As a consequence of the bootstrap equations \cite{GZ} the breather
reflection factors share the structure of the solitonic ones, \cite{gosh}:
\[
R^{(n)}(\eta ,\vartheta ,u)=R_{0}^{(n)}(u)S^{(n)}(\eta ,u)S^{(n)}(i\vartheta ,u)\, \, \, ,\]
where \[
R_{0}^{(n)}(u)=\frac{\left( \frac{1}{2}\right) \left( \frac{n}{2\lambda }+1\right) }{\left( \frac{n}{2\lambda }+\frac{3}{2}\right) }\prod ^{n-1}_{l=1}\frac{\left( \frac{l}{2\lambda }\right) \left( \frac{l}{2\lambda }+1\right) }{\left( \frac{l}{2\lambda }+\frac{3}{2}\right) ^{2}}\quad ;\quad S^{(n)}(x,u)=\prod ^{n-1}_{l=0}\frac{\left( \frac{x}{\lambda \pi }-\frac{1}{2}+\frac{n-2l-1}{2\lambda }\right) }{\left( \frac{x}{\lambda \pi }+\frac{1}{2}+\frac{n-2l-1}{2\lambda }\right) }\, \, \, .\]
 In general \( R_{0}^{(n)} \) describes the boundary independent
properties and the other factors give the boundary dependent ones.

\subsection{Fundamental domain of the parameters}

Since the breather reflection factors can be obtained from the solitonic
ones by the bootstrap principle we concentrate only on \( R(\eta ,\vartheta ,u) \).
From the \[
\sigma (x,u)=\sigma (-x,u)\]
property it follows that it is enough to consider the \( \eta \geq 0\, ,\, \, \vartheta \geq 0 \)
cases. The boundary dependent poles of \( R(\eta ,\vartheta ,u) \)
are due to the factor \( \sigma (\eta ,u) \), whose poles in the
physical strip are located at \begin{equation}
\label{spole}\eqalign{
a._{)}\qquad u =\frac{\eta }{\lambda }-u_{2k+1}-2& \pi m\quad
\textrm{or at }\quad b._{)}\qquad u=\pi -\frac{\eta }{\lambda
}+u_{2k+1}+2\pi m\quad ;\cr k &\geq 0,\ m\geq 0\quad \textrm{both integers}}
\end{equation}
 (The factor \( \sigma (i\vartheta ,u) \) has no pole in the physical
strip). In any Coleman-Thun diagram the incident soliton decays into
a soliton and a breather by the process on diagram \ref{bulk_fusion} (b). Since the
rapidity difference between the particles created is larger than \( \frac{\pi }{2} \),
moreover, in the case of ground state scattering both particles have
to travel towards to wall, we conclude that no Coleman-Thun diagram
exists (except the very special one described on diagram \ref{soliton_diags} (a)). From
this it follows that any of the poles of \( \sigma (\eta ,u) \) in
the physical strip must correspond to boundary bound states, that
is to excited states of the boundary. In a minimal solution we expect
these states to be non degenerate, i.e. they should not appear in
multiplets. For this reason the determinant of the prefactor of \( R(\eta ,\vartheta ,u) \)
must vanish at these poles. The vanishing gives rise to the following
equation 
\[ \sin (\lambda u)=\pm \cos (\eta ) , \] from which we obtain
the location of the zeroes 
\[ u=\frac{\eta }{\lambda }-u_{2k+1}\, ;\, \, \quad k\in \mathbb Z \ .
\]
Since, for general \( \eta  \), among the poles of \( \sigma (\eta ,u) \)
there are some which are not in this set of zeroes, we have to exclude
them by restricting the range of \( \eta  \). This can be achieved
by demanding \( \pi -\frac{\eta }{\lambda }+u_{1}\geq \frac{\pi }{2} \),
or equivalently by\[
0<\eta \leq \frac{\pi }{2}(\lambda +1)=\eta _{0}\, \, \, .\]
 Had we chosen \( \eta  \) outside of this range, which we call the
fundamental range, we would have extra poles in the physical strip
losing the minimality of the solution, thus inducing extra CDD factors
in the reflection amplitude. 

\subsection{Soliton reflection factors on ground state boundary}

\begin{figure}\subfigure[Soliton bulk
pole]{\resizebox*{!}{6cm}{\includegraphics{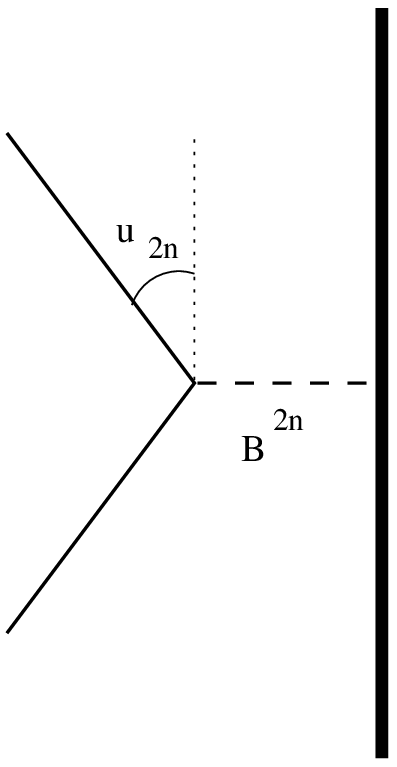}}}
~~~~~~~~~~~~~~~\subfigure[Soliton
decay]{\resizebox*{!}{6cm}{\includegraphics{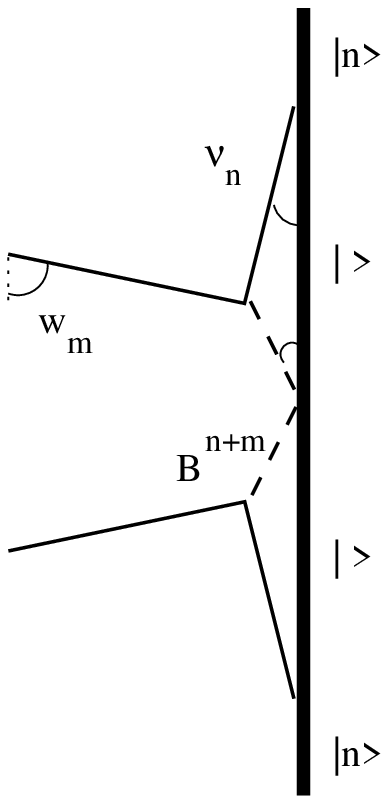}}}
~~~~~~~~~~~~~~~~\subfigure[Crossed
soliton]{\resizebox*{!}{6cm}{\includegraphics{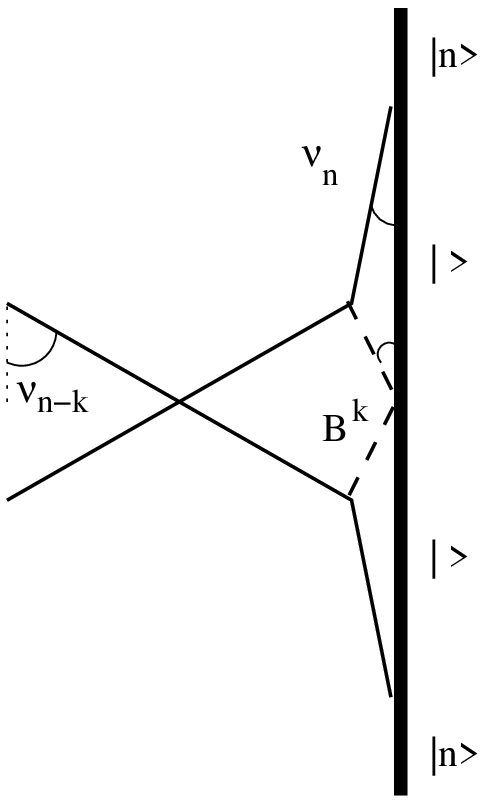}}} ~~~~~
\caption{\label{soliton_diags} Solitonic diagrams}
\end{figure}

Now consider the poles of the soliton reflection factor \( R(\eta ,\vartheta ,u) \).

\begin{itemize}
\item There are boundary independent poles in the physical strip coming
from the factor \( R_{0}(u) \). They are located at \( u_{n}\, ,\, n=1,2,\dots  \),
and can be described by diagram \ref{soliton_diags} (a). Clearly the diagram does not
exist for \( Q(\eta ,\vartheta ,u) \) and the prefactor \( Q_{0}(u) \)
takes care of this. 
\item The boundary dependent poles are located at\[
\nu _{n}=\frac{\eta }{\lambda }-u_{2n+1}\quad ;\qquad n=0,1,\dots \, \, \, ,\]
where the upper limit for \( n \) can be determined by restricting
\( \nu _{n} \) to the physical region. To each of the poles above
we associate a boundary bound state denoted by \( |n\rangle  \) with
energy \[
m_{|n\rangle }-m_{|\, \rangle }=M\cos (\nu _{n})\, \, \, .\]
 (\( M \) is the soliton mass). Clearly the state \( |n\rangle  \)
is present in all reflection factors \( P^{\pm }(\eta ,\vartheta ,u),\, Q(\eta ,\vartheta ,u) \). 
\end{itemize}
In the generic case the reflection factor \( R(\eta ,\vartheta ,u) \)
has no zero in the physical strip. The special cases will be discussed
separately.

\subsection{Solitonic reflection factors on excited boundary states}

\begin{figure}
\begin{center}
\resizebox*{!}{6cm}{\includegraphics{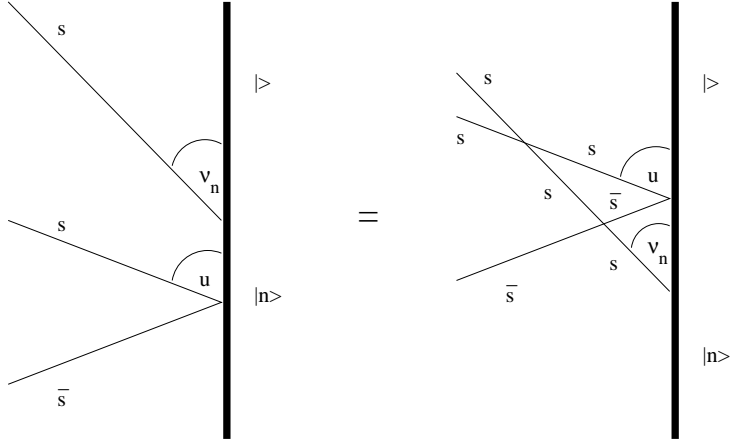}}
\end{center}
\caption{\label{qbootstrap} The bootstrap equation for the amplitude $Q$.}
\end{figure}

The reflection factor \( R_{|n\rangle }(\eta ,\vartheta ,u) \) on
the boundary bound state \( |n\rangle  \) can be computed from the
boundary bootstrap equations, the simplest of them is (see Fig. \ref{qbootstrap})
\begin{equation}
\label{qbst}
Q_{|n\rangle }(\eta ,\vartheta ,u)=a(u-\nu _{n})Q(\eta ,\vartheta ,u)b(u+\nu _{n})\, \, \, .
\end{equation}
 Using (\ref{abc}) and the relation\[
a(u+\nu _{0})a(u-\nu _{0})=\frac{\cos (\eta )\cos (\lambda u-\bar{\eta })\sigma (\bar{\eta },u)}{\cos (\bar{\eta })\cos (\lambda u+\eta )\sigma (\eta ,u)}\quad ;\qquad \bar{\eta }=2\eta _{0}-\eta \]
 we obtain
\begin{equation}\label{Qnexpl}
Q_{|n\rangle }(\eta ,\vartheta ,u)=Q(\bar{\eta },\vartheta
,u)a_{n}(\eta ,u)\, \, \, ,
\end{equation}
 where \[
a_{n}(\eta ,u)=\frac{a(u+\nu _{n})a(u-\nu _{n})}{a(u+\nu _{0})a(u-\nu _{0})}=\prod _{l=1}^{n}\left\{ 2\left( \frac{\eta }{\pi }-l\right) \right\} \, \, \, .\]
The matrix structure of \( R_{|n\rangle }(\eta ,\vartheta ,u) \)
follows from the boundary Yang-Baxter equation, and is identical to
that of in (\ref{Rsas}) modulo a possible soliton antisoliton interchange.
Performing the bootstrap calculation shows that the interchange really
takes place and we have
\begin{equation}
P^{\pm }_{|n\rangle }(\eta ,\vartheta ,u)=P^{\mp }(\bar{\eta
},\vartheta ,u)a_{n}(\eta ,u)\, \, \, .\end{equation} 
Making a comparison between \( R_{|0\rangle }(\eta ,\vartheta ,u) \)
and \( R(\eta ,\vartheta ,u) \) we observe that they are related
by the transformations, \( \eta \leftrightarrow \bar{\eta }\, ,\, \, s\leftrightarrow \bar{s} \).
Consequently these transformations change the roles of the two lowest
lying boundary states, namely $|\ \rangle$ and $|0\rangle$. For this reason we conjecture that they are
the quantum manifestation of the classical \( \Phi \leftrightarrow \frac{2\pi }{\beta }-\Phi  \)
transformation. Clearly the energy of the vacuum state, $|\ \rangle$, is not determined by the boostrap.
In the light of the preceeding remark, however,  we expect it to satisfy 
\[ 
m_{|0\rangle}(\eta,\vartheta)=m_{|\, \rangle }(\eta,\vartheta)+M\cos (\nu _{0})=m_{|\, \rangle }(\bar \eta,\vartheta)
\, \, . 
\] 
Let us turn to the analysis of the pole structure of \( R_{|n\rangle
}(\eta ,\vartheta ,u) \).  Since \( \eta _{0}<\bar{\eta }<2\eta _{0}
\), the poles of \( \sigma (\bar{\eta },u) \) which are of the form
(\ref{spole}.a) are located at\[ w_{k}=\frac{\bar{\eta }}{\lambda
}-u_{2k+1}=\pi -\frac{\eta }{\lambda }-u_{2k-1}\quad ;\qquad
k=0,1,\dots \, \, \, \, ,\] while those of the form of (\ref{spole}.b)
are located at \( \nu _{k}\, ,\, \, k=0,-1,\dots \).  The factor \(
a_{n}(\eta ,u) \) has simple poles at \( \nu _{0} \) and \( \nu _{n}
\), and double poles at \( \nu _{k}\, ,\, \, k=1,2,\dots ,n-1 \).  The
\( \nu \)-type poles can be explained by diagram \ref{soliton_diags}
(c) (which gives a second order pole) except for \( \nu _{n} \) where
the breather line is absent and the diagram reduces to a crossed
channel soliton emission-absorption process (which has order $1$).
Diagram \ref{soliton_diags} (b) explains the poles \( w_{m} \) but
only for \( w_{m}\geq \nu _{n} \). Naively, the diagram gives a second
order pole but in this case the breather reflection in the middle is
at an angle where the reflection factor has a simple zero, so the
order of the diagram is reduced to one.  For \( w_{m}<\nu _{n} \) we
have a boundary bound state, which is denoted by \( |n,m\rangle
\). The computation of the reflection factors on this state is
completely analogous to eqn. (\ref{qbst}), by replacing \( \nu_n \)
and $Q$ by \( w_m\) and \( Q_{\vert n \rangle }\). Indeed
\[\eqalign{
Q_{|n, m\rangle }(\eta ,\vartheta ,u)&=a(u-w_{m})Q_{\vert n\rangle
}(\eta ,\vartheta ,u)b(u+w_{m})\cr
& =a(u-w_{m})Q(\bar{\eta },\vartheta ,u)b(u+w_{m})a_n(\eta ,u)\cr
& =Q_{\vert m\rangle }(\bar{\eta },\vartheta ,u)a_n(\eta ,u) .}
\]
(In writing the third equality we used eq. (\ref{qbst}) and \( w_m=\nu_m(\bar{\eta})\)). 
As a result, using (\ref{Qnexpl}), we
have:
\begin{equation}
Q_{|n,m\rangle }(\eta ,\vartheta ,u)=Q(\eta ,\vartheta ,u)a_{n}(\eta
,u)a_{m}(\bar{\eta },u)\, \, \, .\end{equation}  
In a similar way we obtain
\begin{equation}\label{utolso}
P^{\pm }_{|n,m\rangle }(\eta ,\vartheta ,u)=P^{\pm }(\eta ,\vartheta
,u)a_{n}(\eta ,u)a_{m}(\bar{\eta },u)\, \, \, .\end{equation} 
 The situation concerning the poles of \( R_{|n,m\rangle }(\eta ,\vartheta ,u) \)
is the same as in the previous case if we make the \( \eta \leftrightarrow \bar{\eta } \)
replacement. Thus the poles, which {\it can not} be explained by
diagrams, (and so describe new bound states), 
are located at \( \nu _{k} \) for \( \nu _{k}<w_{m} \). The
corresponding bound state is denoted as \(\vert n,m,k\rangle \).  
Now it is easy to see how eq.(\ref{Qnexpl} -\ref{utolso}) and the generation
of new boundary states
would go on so we could turn to the analysis of
the general case, but before doing this we analyse the breather sector.

\subsection{Breather reflection factors on ground state boundary}

The boundary independent poles of the reflection factor \( R^{(n)}(\eta ,\vartheta ,u) \)
come from \( R^{(n)}_{0}(u) \). In the physical strip \( R^{(n)}_{0} \)
has simple pole 

\begin{itemize}
\item at \( \frac{\pi }{2} \), which corresponds to the emission of a zero
momentum breather, 
\item at \( u_{k} \) for \( k=1,2,\dots ,n-1 \), they can be explained
in terms of diagram \ref{breather_diag1} (a),
\item at \( \frac{\pi }{2}-u_{n} \), which is related to the breather version
of diagram \ref{soliton_diags} (a) by forming \( B^{2n} \) or if this is not in the spectrum
then the soliton version of diagram \ref{breather_diag1} (a). 
\item The double poles at \( \frac{\pi }{2}-u_{k} \), for \( k=1,2,\dots ,n-1 \)
can be explained by diagram \ref{breather_diag1} (b).  \vspace{0.3cm}

\end{itemize}
\vspace{0.3cm}
{\centering \begin{figure}~~~~~~~~~~~~~\subfigure[Breather
triangle]{\resizebox*{!}{5cm}{\includegraphics{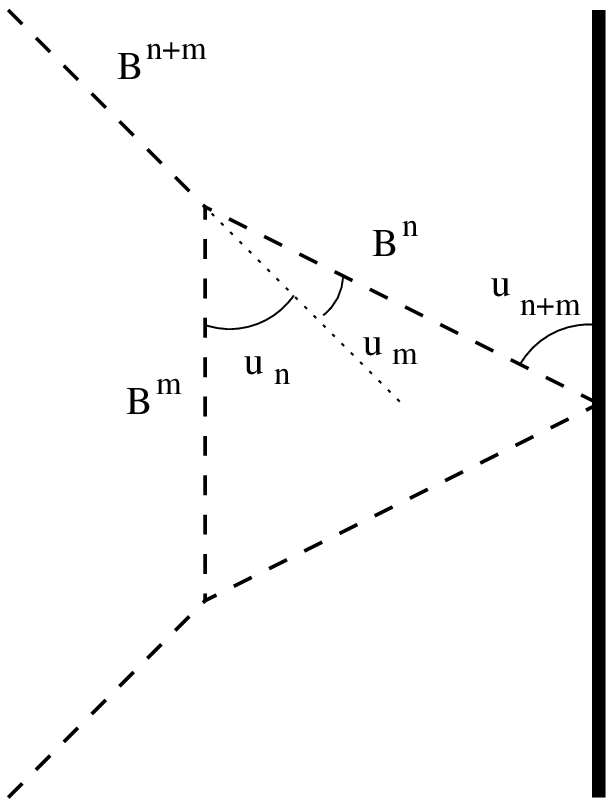}}}
~~~~~~~~~~~~~~~~\subfigure[Breather
poles]{\resizebox*{!}{5cm}{\includegraphics{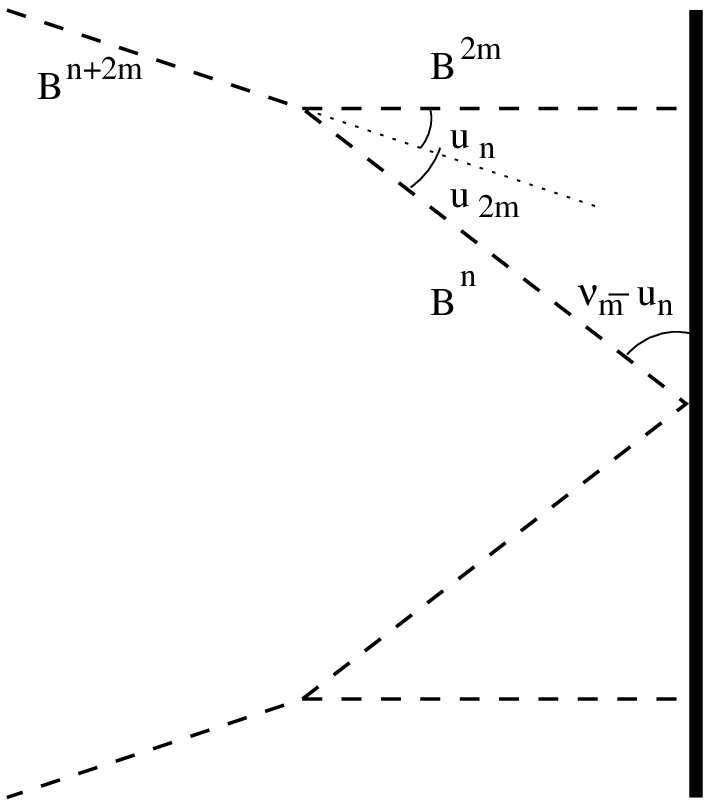}}} ~~~~~~~~~~
\caption{\label{breather_diag1} Breather diagrams I}
\end{figure}\par}
\vspace{0.3cm}

The boundary dependent poles are located at\[
u=\frac{1}{2}(\nu _{k}-w_{n-k})=\frac{\eta }{\lambda }-\frac{\pi }{2}+u_{n-2k-1}\quad ;\qquad k=0,\dots ,\left[ \frac{n-1}{2}\right] \, \, \, ,\]
 If these poles are in the physical strip then they correspond to
the creation of the state \( |k,n-k\rangle  \). To see this note
that \( |k,n-k\rangle  \) can be created in two ways (Fig. \ref{shifting}): either first
a soliton (say) reflecting at \( \nu _{k} \) on \( |\, \rangle  \)
generating \( |k\rangle  \), followed by an antisoliton reflecting
at \( w_{n-k}<\nu _{k} \) on \( |k\rangle  \), or, moving the antisoliton
trajectory upwards, first the antisoliton reflecting at \( w_{n-k} \)
on \( |\, \rangle  \) (without creating any new states), then the
reflected antisoliton fusing with the soliton forming \( B^{n} \)
with rapidity \( \frac{1}{2}(\nu _{k}-w_{n-k}) \), which is finally
creating \( |k,n-k\rangle  \). This mechanism was first described in
\cite{patr}.

Finally we remark that the reflection factor also has zeroes
at \[
u=\pm \frac{\eta }{\lambda }+\frac{\pi }{2}+u_{n-2k-1}\quad ;\qquad k=0,\dots ,n-1\, \, \, .\]

{\centering \begin{figure}~~~~~~~~~~~~~\subfigure{\resizebox*{!}{6cm}{\includegraphics{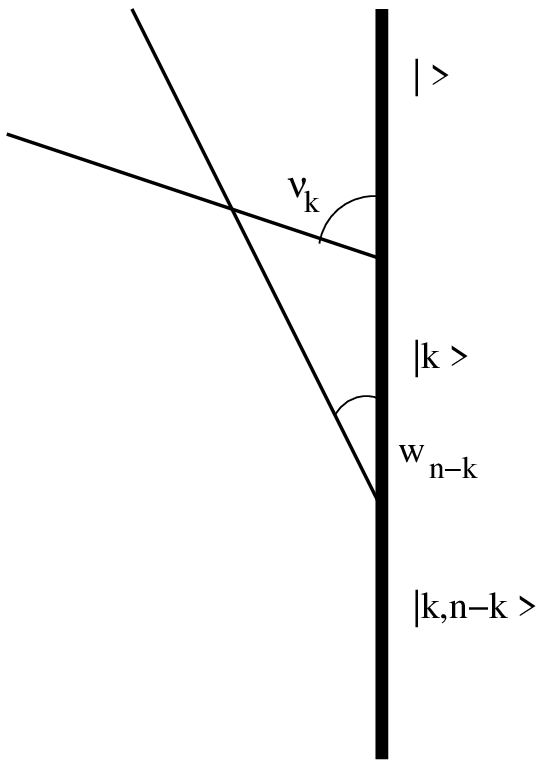}}}
~~~~~~~~~~~~~~~~\subfigure{\resizebox*{!}{6cm}{\includegraphics{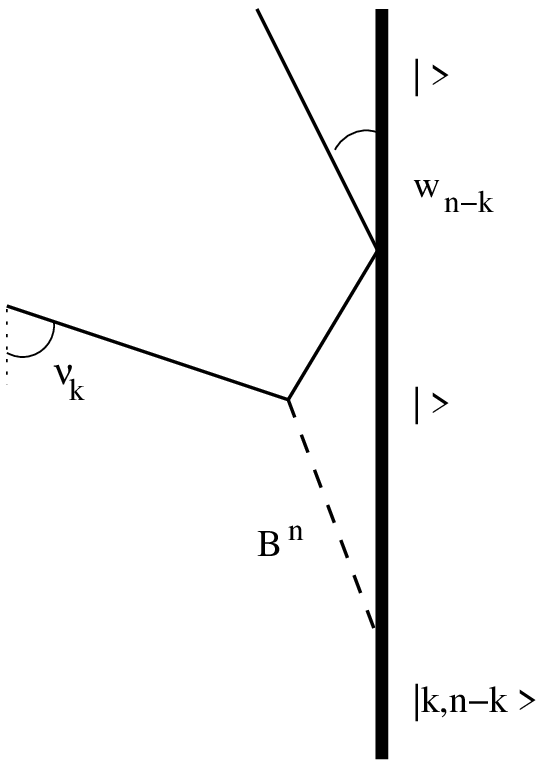}}} ~~~~~~~~~~
\caption{\label{shifting} Breather pole explanations related by
shifting soliton trajectories}
\end{figure}\par}

\subsection{Breather reflection factors on excited boundary states}

Breathers can emerge as a virtual fusion of a soliton and antisoliton
as shown on diagram \ref{bulk_fusion} (b). Since this fusion can take
place not only before the soliton and antisoliton reflect on the
boundary but also after their reflection, a bootstrap equation can be
obtained for the breather reflection factors on excited boundary
states \cite{GZ, FK}, graphically it is shown on
Fig. \ref{bbootstrap}.

{\centering \begin{figure}~~~~~~~~~~~~~~~~~~\subfigure{\resizebox*{!}{6cm}{\includegraphics{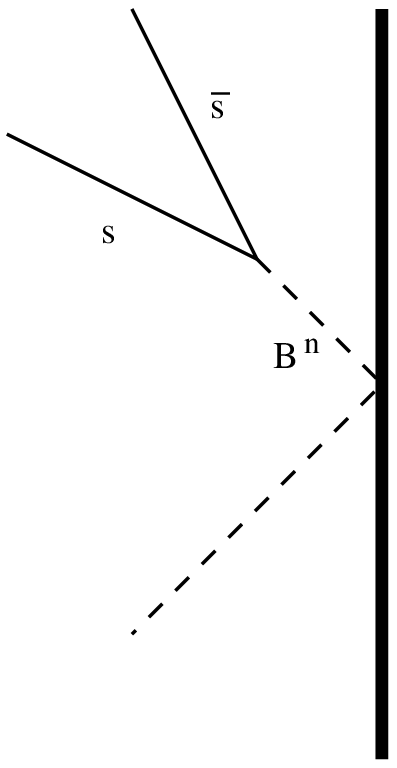}}}
~~~~~~~~~~~~~~~~~~~~~~~~~~~~~~~\subfigure{\resizebox*{!}{6cm}{\includegraphics{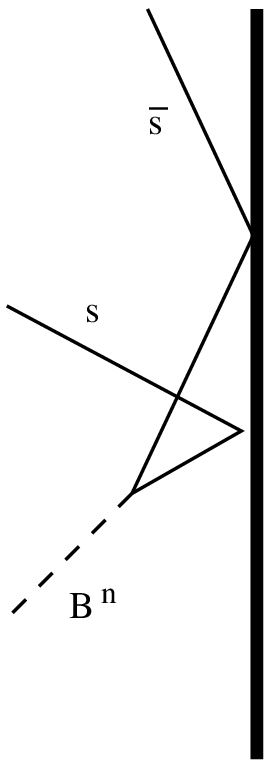}}} ~~~~~~~~~~
\caption{\label{bbootstrap} Breather reflection factors from bootstrap}
\end{figure}\par}

Clearly this reflection factor has the same structure as the solitonic
one:\begin{equation}
\label{b1ref}
R^{(n)}_{|k\rangle }(\eta ,\vartheta ,u)=R^{(n)}(\bar{\eta },\vartheta ,u)b_{k}^{n}(\eta ,u)\, \, \, ,
\end{equation}
 where \[\eqalign{
b_{k}^{n}(\eta ,u)&=a_{k}(u+\frac{\pi }{2}-u_{n})a_{k}(u-\frac{\pi
}{2}+u_{n})
\cr &=\prod _{l=1}^{\min (n,k)}\left\{ \frac{2\eta }{\pi }-\lambda
+n-2l\right\} \left\{ \frac{2\eta }{\pi }+\lambda -n-2(k+1-l)\right\} ,} \, \, \, \]
and a cancellation between the various factors has also been taken
into account. Let us focus on the pole structure. The \( \eta  \)-independent
poles of \( R^{(n)}_{|k\rangle }(\eta ,\vartheta ,u) \) are located
at the same place as those of \( R^{(n)}(\eta ,\vartheta ,u) \) and
have the same explanation. 

The \( \eta  \)-dependent poles of 
\(R^{(n)}_{|k\rangle }(\eta ,\vartheta ,u)  \)
are considered in three steps. First consider the following expression\[
\prod _{l=1}^{\min (n,k)}\left\{ \frac{2\eta }{\pi }+\lambda -n-2(k+1-l)\right\} \, \, \, ,\]
 which has poles at \begin{equation}
\label{wnupoles}
u=\frac{1}{2}(w_{l-k}-\nu _{n+k-l})=\frac{\pi }{2}-\frac{\eta }{\lambda }+u_{n+2(k-l)+1}
\end{equation}
\begin{equation}
\label{plpoles}
u=\frac{\pi }{2}+\frac{\eta }{\lambda }-u_{n+2(k-l)+1}\, \, \, ,
\end{equation}
and for each pole at \( u \) it has a zero at \( -u \). The poles/zeroes
for \( l=0 \) and \( l=\min (n,k) \) are simple, while for the other
\( l \) -s are double. The simple pole at \( l=0 \) for (\ref{wnupoles})
describes the \( |k\rangle \mapsto |k+n\rangle  \) boundary bound
state changing process. The other simple pole at \( l=n \) for \( k\geq n \)
can be explained by the crossed version of this process, while for
\( k<n \) we have diagram \ref{breather_diag2} (a). This diagram also explains all the
double poles in (\ref{wnupoles}). In the case of poles given by eqn.
(\ref{plpoles}) we have diagram \ref{breather_diag2} (b). (Here for \( l=k=0 \) the diagram
simplifies, the initial breather does not decay in the bulk, just
fuses with the soliton). On diagram \ref{breather_diag2} (b) the soliton reflects on the
wall with rapidity \( -\nu _{l+k} \). It has no zero here, however,
summing up all possible diagrams ensures that the order of the diagram
is one in the \( l=k=0 \) case and two otherwise. 

\vspace{0.3cm}
{\centering \begin{figure}~~~\subfigure[Breather
direct]{\resizebox*{!}{6cm}{\includegraphics{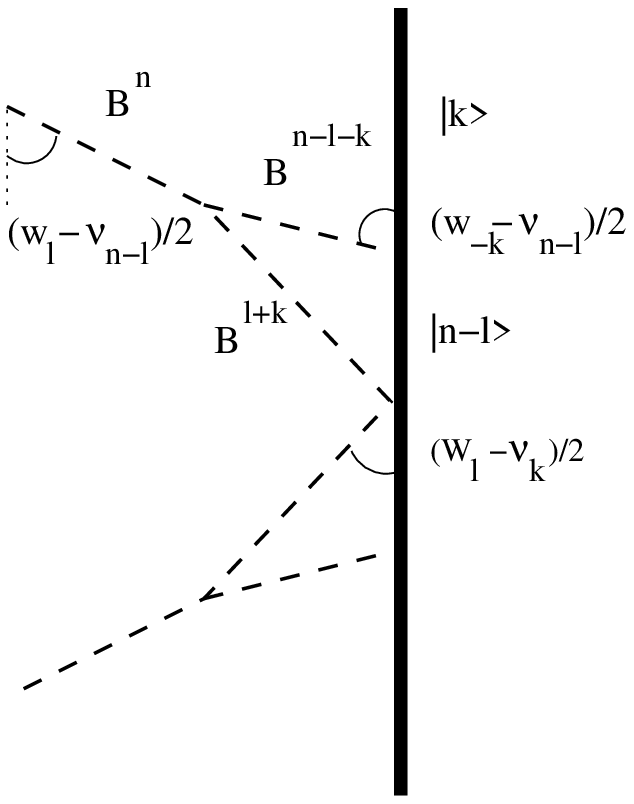}}}
~~\subfigure[Breather
crossed]{\resizebox*{!}{6cm}{\includegraphics{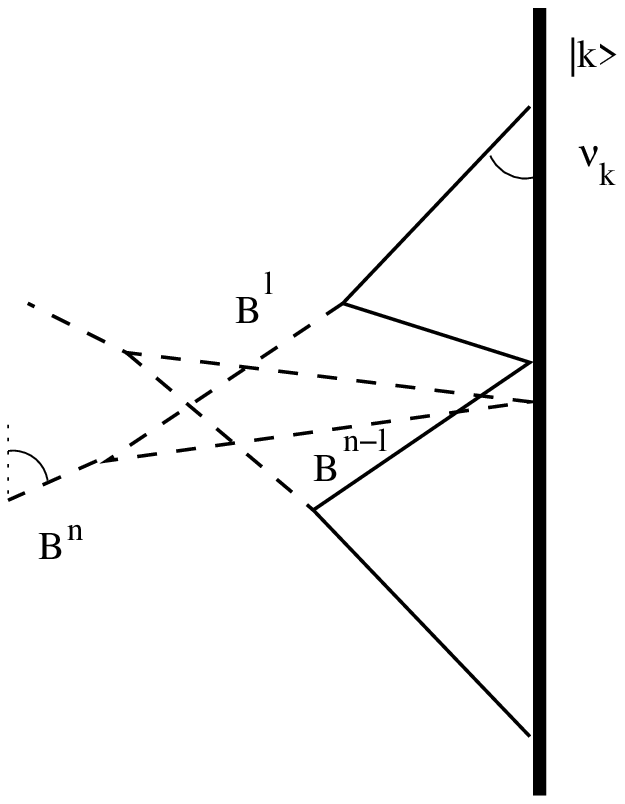}}}
~~~~~~\subfigure[Breather
decay]{\resizebox*{!}{6cm}{\includegraphics{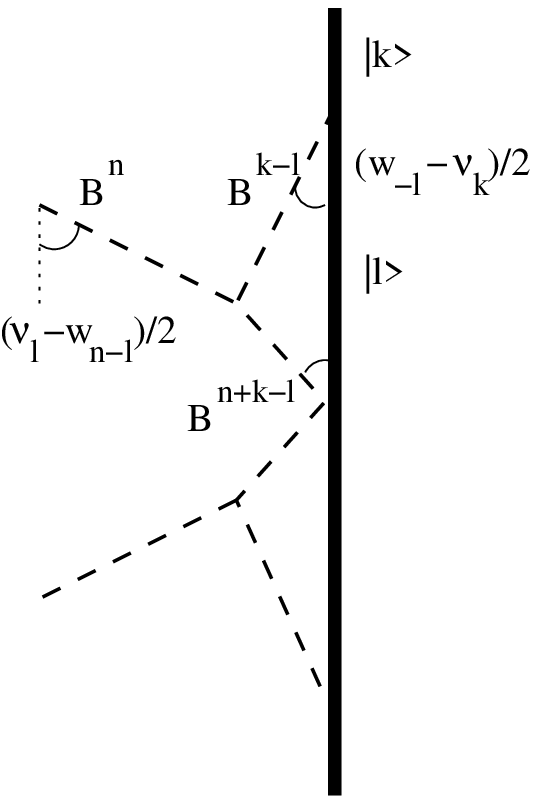}}} 
\caption{\label{breather_diag2} Breather diagrams II}
\end{figure}\par}
\vspace{0.3cm}

Now consider the remaining \( \eta  \)-dependent factors of
 \( R^{(n)}_{|k\rangle }(\eta ,\vartheta ,u) \):
\begin{equation}
\label{fkk2}
S^{(n)}(\bar{\eta },u)\prod ^{\min (n,k)}_{l=1}\left\{ \frac{2\eta
}{\pi }-\lambda +n-2l\right\} =\prod ^{k-1}_{l=0}\frac{\left(
\frac{\eta }{\lambda \pi }-\frac{1}{2}+\frac{n-2l-1}{2\lambda }\right)
}{\left( \frac{\eta }{\lambda \pi }+\frac{1}{2}+\frac{n-2l-1}{2\lambda
}\right) }\prod ^{n}_{l=k+1}\frac{\left( \frac{\eta }{\lambda \pi
}+\frac{1}{2}+\frac{n-2l-1}{2\lambda }\right) }{\left( \frac{\eta
}{\lambda \pi }-\frac{1}{2}+\frac{n-2l+1}{2\lambda }\right) }\ .
\end{equation}
For \( k>n \) the second product on the r.h.s. disappears and the first product
contains \( n \) factors giving exactly \( S^{(n)}(\eta ,u) \).
In the \( k=0 \) case the first product disappears and the second
gives \( S^{(n)}(\bar{\eta },u) \). From these two limiting cases
we can understand the generic case. 

Now we explain the poles of the first product on the r.h.s. of
(\ref{fkk2}), 
which is nothing but
the first \( k \) factors of \( S^{(n)}(\eta ,u) \). On the ground
state boundary its poles at \[
u=\frac{1}{2}(\nu _{l}-w_{n-l})=\frac{\eta }{\lambda }-\frac{\pi }{2}+u_{n-2l+1}\quad ;\qquad l=0,\dots ,k-1\]
would correspond to the creation of the state \( |l,n-l\rangle  \).
Now, however, they correspond to the creation of the state \( |l,n-l,k\rangle  \).
In order for this state to exist we need \( l<k \), (this is clearly
satisfied), and \( w_{n-l}>\nu _{k} \). Alternatively, it is not
hard to see that if \( w_{n-l}<\nu _{k} \) then we have diagram
\ref{breather_diag2} (c),
where under this condition \( B^{n+k-l} \) travels towards the wall. 

The second product on the r.h.s. of (\ref{fkk2}) has simple poles at
(\ref{wnupoles}, \ref{plpoles})
for \( l=k,\dots ,n-1 \). The poles of the form (\ref{wnupoles})
correspond to the creation of the state \( |k,l-k,n-l+k\rangle  \).
If this state is not in the spectrum, that is if \( \nu _{k}<w_{l-k} \)
then we have diagram \ref{breather_diag2} (a), where \( B^{l+k} \) travels towards the
wall. For the poles of type (\ref{plpoles}) we have diagram
\ref{breather_diag2} (b). 

For the state \( |k,m\rangle  \) the bootstrap equation giving (\ref{b1ref})
generalizes to 
\[
R^{(n)}_{|k,m\rangle }(\eta ,\vartheta ,u)=R^{(n)}(\eta ,\vartheta ,u)b_{k}^{n}(\eta ,u)b_{m}^{n}(\bar{\eta },u)\, \, \, ,\]
and from this the general expression can be conjectured. 

The upshot of this investigation of the pole structure of the breather
reflection factors is the realization that the breathers do not create
new type of boundary bound states, they merely give an alternative
way of jumping between the ones generated by the solitons.

\subsection{The general spectrum and the associated reflection factors}

From the previous considerations it is clear how to generalize the
results. The spectrum of boundary bound states can be parametrized
by a sequence of integers \( |n_{1},n_{2},\dots ,n_{k}\rangle  \),
whenever the 
\[ \frac{\pi }{2}\geq \nu _{n_{1}}>w_{n_{2}}>\nu _{n_{3}}>\dots \geq 0 
\]
condition holds. The mass of such a state is 
\[
m_{|n_{1},n_{2},\dots ,n_{k}\rangle }=m_{|\, \rangle }(\eta,\theta)+M\sum _{i\textrm{ odd}}\cos (\nu _{n_{i}})+M\sum _{i\textrm{ even}}\cos (w_{n_{i}})\, \, \, .\]
The reflection factors depend on which sectors we are considering.
In the even sector, i.e. when \( k \) is even, we have\[
Q_{|n_{1},n_{2},\dots ,n_{k}\rangle }(\eta ,\vartheta ,u)=Q(\eta ,\vartheta ,u)\prod _{i\textrm{ odd}}a_{n_{i}}(\eta ,u)\prod _{i\textrm{ even}}a_{n_{i}}(\bar{\eta },u)\, \, \, ,\]
and \[
P^{\pm }_{|n_{1},n_{2},\dots ,n_{k}\rangle }(\eta ,\vartheta ,u)=P^{\pm }(\eta ,\vartheta ,u)\prod _{i\textrm{ odd}}a_{n_{i}}(\eta ,u)\prod _{i\textrm{ even}}a_{n_{i}}(\bar{\eta },u)\, \, \, ,\]
for the solitonic processes and \[
R^{(n)}_{|n_{1},n_{2},\dots ,n_{k}\rangle }(\eta ,\vartheta ,u)=R^{(n)}(\eta ,\vartheta ,u)\prod _{i\textrm{ odd}}b^{n}_{n_{i}}(\eta ,u)\prod _{i\textrm{ even}}b^{n}_{n_{i}}(\bar{\eta },u)\, \, \, \]
for the breather process. In the odd sector, i.e. when \( k \) is
odd, the same formulae apply if in the ground state reflection factors
the \( \eta \leftrightarrow \bar{\eta } \) and \( s\leftrightarrow \bar{s} \)
changes are made. 

In order to prove the existence of these states (i.e. the absence
of Coleman-Thun diagrams) the proof from \cite{patr} can be used
in the general case, (see the remark at the end of the next subsection),
while a straightforward modification of the proof given for the Neumann
boundary \cite{saj} applies in the \( \eta =\eta _{0} \) case. 

These boundary states have the following creation/annihilation rules:

\bigskip{}
{\centering \begin{tabular}{|c|c|c|c|}
\hline 
Initial state&
particle&
rapidity&
final state\\
\hline
\hline 
\( |n_{1},\dots ,n_{2k}\rangle  \)&
\( s,\bar{s} \)&
\( \nu _{n} \)&
\( |n_{1},\dots ,n_{2k},n\rangle  \)\\
\hline 
\( |n_{1},\dots ,n_{2k-1}\rangle  \)&
\( s,\bar{s} \)&
\( w_{n} \)&
\( |n_{1},\dots ,n_{2k-1},n\rangle  \)\\
\hline 
\( |n_{1},\dots ,n_{2k},n_{2k+1},\dots \rangle  \)&
\( B^{n} \)&
\( \frac{1}{2}(\nu _{l}-w_{n-l}) \)&
\( |n_{1},\dots ,n_{2k},l,n-l,n_{2k+1},\dots \rangle  \)\\
\hline 
\( |n_{1},\dots ,n_{2k-1},n_{2k},\dots \rangle  \)&
\( B^{n} \)&
\( \frac{1}{2}(w_{l}-\nu _{n-l}) \)&
\( |n_{1},\dots ,n_{2k-1},l,n-l,n_{2k},\dots \rangle  \)\\
\hline 
\( |n_{1},\dots ,n_{2k},\dots \rangle  \)&
\( B^{n} \)&
\( \frac{1}{2}(\nu _{-n_{2k}}-w_{n+n_{2k}}) \)&
\( |n_{1},\dots ,n_{2k}+n,\dots \rangle  \)\\
\hline 
\( |n_{1},\dots ,n_{2k-1},\dots \rangle  \)&
\( B^{n} \)&
\( \frac{1}{2}(w_{-n_{2k-1}}-\nu _{n+n_{2k-1}}) \)&
\( |n_{1},\dots ,n_{2k-1}+n,\dots \rangle  \)\\
\hline
\end{tabular}\par}
\bigskip{}

The other poles in the reflection factors can be explained exactly
in the same way as in the Dirichlet or in the Neumann cases \cite{patr, saj}.

\subsection{Special cases}

Let us consider the \( \eta =\eta _{0} \) case. Since now \( \eta =\bar{\eta } \)
the two types of poles coincide: \( \nu _{n}=w_{n}=\frac{\pi }{2}-u_{2n} \)
thus in the labeling of any state \( |n_{1},n_{2},\dots ,n_{k}\rangle  \)
we have monotonically increasing sequences of non-negative integer
numbers \( n_{i+1}>n_{i} \). Comparing the state \( |\, \rangle  \)
to \( |0\rangle  \) we observe that they have the same energy, which
is the lowest in the entire spectrum, moreover the reflection factors
on them coincide for the breathers and for \( Q \). These states
are not the same in general, however, since for the solitonic reflection
factors we have the relation\begin{equation}
\label{sbars}
P^{\pm }(\eta _{0},\vartheta ,u)=P_{|0\rangle }^{\mp }(\eta _{0},\vartheta ,u)\, \, \, .
\end{equation}
Thus these two states give two inequivalent vacua of the theory connected
by the \( s\leftrightarrow \bar{s} \) transformation, which symmetry
is then spontaneously broken. Emphasizing this we introduce the following
notation \( |0\rangle _{+}=|\, \rangle  \) and \( |0\rangle _{-}=|0\rangle  \).
We can also relabel all the states referring to the two sectors as
\[\eqalign{
|n_{1},\dots ,n_{k}\rangle _{+}&=\left\{ \begin{array}{c}
|n_{1},\dots ,n_{k}\rangle \, \, \, \,  \textrm{if}\, \, \,  k\textrm{ is even}\\
|0,n_{1},\dots ,n_{k}\rangle \, \, \, \,  \textrm{if}\, \, \,  k\textrm{ is odd}
\end{array}\right. \ ;\cr |n_{1},\dots ,n_{k}\rangle _{-}&=\left\{ \begin{array}{c}
|n_{1},\dots ,n_{k}\rangle \, \, \, \,  \textrm{if}\, \, \,  k\textrm{ is odd}\\
|0,n_{1},\dots ,n_{k}\rangle \, \, \, \,  \textrm{if}\, \, \,  k\textrm{ is even}
\end{array}\right. ;}\, \, \, \, \]
where now \( n_{i}>0\, ,\, \, \forall i \) hold. The energies of
the states \( |n_{1},\dots ,n_{k}\rangle _{+} \) and \( |n_{1},\dots ,n_{k}\rangle _{-} \)
are equal, moreover the soliton/antisoliton reflection factors on
them are related by the \( s\leftrightarrow \bar{s} \) transformation
similarly to eqn.(\ref{sbars}). The breathers create states only
within the sectors, while the solitons (antisolitons) jump between
the two sectors. It is interesting that \( B^{n} \) creates the state
\( |l,n-l\rangle _{+} \) when it reflects on \( |0\rangle _{+} \)
with rapidity \( u_{n-2l} \), for which the process on diagram
\ref{breather_diag1} (a)
is also present, thus boundary bound state creation and Coleman-Thun
diagrams coexist, just like in the case of Neumann boundary condition
\cite{saj}. 

The case of the Neumann boundary condition can be obtained from the
one investigated above by taking the \( \vartheta \to 0 \) limit.
In this case the \( s\leftrightarrow \bar{s} \) symmetry is not broken
(\( P^{+}(\eta _{0},0,u)=P^{-}(\eta _{0},0,u) \)), and we have just
one ground state and one sector. 

The \( \Phi \leftrightarrow -\Phi  \)
symmetric boundary conditions are realized either by \( \eta \equiv 0 \) 
(with \(\vartheta \) running in its fundamental domain) or by \( \vartheta \equiv 0 \) 
(with \(\eta \) in its fundamental domain).
In these models the spectrum is the same as in the general case, the
only difference is that \( P^{+} \) and \( P^{-} \) coincide. 

The case of the Dirichlet boundary condition, which was analysed in
detail in \cite{patr}, can be obtained by taking the \( \vartheta \to \infty  \)
limit. As a result \( Q \) vanishes, terms containing functions of
\( \vartheta  \) disappear and we have \[
P^{\pm }(\eta ,u)=\cos (\eta \pm \lambda u)R_{0}(u)\frac{\sigma (\eta ,u)}{\cos (\eta )}\, \, \, .\]
The consequence of the prefactor \( \cos (\eta \pm \lambda u) \)
is that now on even walls only solitons can create states while on
odd ones only antisolitons. The spectrum of boundary bound states
and their masses are exactly the same as in the general case. Since
the breather reflection factors have the same pole structure as in
the general case, the proof given in \cite{patr} for the existence
of boundary bound states, and the explanation of the poles of the
reflection factors can be adopted straightforwardly to the general
case considered in this paper.

\section{The UV-IR correspondence}

Now we would like to figure out the assignment between the IR parameters
\( (\eta ,\vartheta ) \) and the UV parameters \( (\phi _{0},M_{0}) \).
The fundamental range of the IR parameters is \( 0\leq \eta \leq \eta _{0}\, ,\, \, \vartheta \geq 0 \),
while that of UV ones is \( 0\leq \phi _{0}\leq \frac{\pi }{\beta }\, ,\, \, M_{0}\geq 0 \).
We suppose the absence of anomalies thus the classical symmetries
survive at the quantum level. In the first step we consider the special
cases, which are at the boundaries of the fundamental domains. 

The spontaneously broken \( \mathbb Z_{2} \) symmetric cases can
be described by the lines \( \eta =\eta _{0} \) in the IR case and
\( \phi _{0}=\frac{\pi }{\beta } \) in the UV one. The Neumann end
of these lines corresponds to \( \vartheta =0 \) and \( M_{0}=0 \),
respectively. We also have the unbroken \( \mathbb Z_{2} \) symmetric
line \( \phi _{0}=0 \), \( 0\leq M_{0}\leq \infty  \). In the domain
of the IR parameters it may be realized in two parts in accord with
the classical behavior in eq.(\ref{mcritcl}): on the \( \vartheta =0 \)
line \( \eta  \) is decreasing from its maximal value \( \eta _{0} \)
(i.e. from the Neumann point) to zero (\( M_{0}<M_{crit} \)), while
on the \( \eta =0 \) line \( \vartheta  \) is increasing from zero
to infinity (Dirichlet) when \( M_{0}>M_{crit} \). (\( M_{crit} \)
appearing here may depend on the sine-Gordon coupling \( M_{crit}=M_{crit}(\lambda ) \),
and this dependence is determined by our considerations only in the classical case 
$M_{crit}(classical)={{4m}\over{\beta^2}}$). In the
Dirichlet case, which corresponds to taking the \( \vartheta \rightarrow \infty  \)
or the \( M_{0}\rightarrow \infty  \) limit, the exact relation between
the remaining parameters is known \cite{GZ}: 
\[ 
2\eta =(\lambda +1)\beta \phi _{0} \, \, .
\] 
Since the boundary potential is periodic in \( \phi _{0} \) with
period \( \frac{4\pi }{\beta } \) we conjecture that the UV-IR relation
contains only functions of the form of \( \cos (n\beta \phi _{0}/2) \)
or \( \sin (n\beta \phi _{0}/2) \), or in the simplest case only
terms with \( n=1 \). We have already seen that the \( \eta \leftrightarrow \bar{\eta }\, ,\, \, s\leftrightarrow \bar{s} \)
transformation corresponds to \( \Phi \leftrightarrow \frac{2\pi }{\beta }-\Phi  \),
that is at the level of the Lagrangian either to \( M_{0}\leftrightarrow -M_{0} \)
or to \( \phi _{0}\leftrightarrow \frac{2\pi }{\beta }-\phi _{0} \).
Combining these transformation properties with the Dirichlet limit,
where \( \eta  \) is linearly related to \( \phi _{0} \), we expect
that in the UV-IR relation either \( \cos (\frac{\eta }{\lambda +1}) \)
or \( \sin (\frac{\eta }{\lambda +1}) \) appear. At the IR level
we also would like to respect the \( \eta \leftrightarrow i\vartheta  \)
symmetry of the ground state reflection factors. Collecting all the
required properties and restricting to the simplest possible case
we are lead to the following formula: \begin{eqnarray*}
\cos \left( \frac{\eta }{\lambda +1}\right) \cosh \left( \frac{\vartheta }{\lambda +1}\right)  & = & \frac{M_{0}}{M_{crit}}\cos \left( \frac{\beta \phi _{0}}{2}\right) \\
\sin \left( \frac{\eta }{\lambda +1}\right) \sinh \left( \frac{\vartheta }{\lambda +1}\right)  & = & \frac{M_{0}}{M_{crit}}\sin \left( \frac{\beta \phi _{0}}{2}\right) \, \, \, .
\end{eqnarray*}
It is not difficult to check that these relations define a mapping
between the fundamental domains of \( (\eta ,\vartheta ) \) and of \(
(\phi _{0},M_{0}) \), which respects all the special cases. Note that
these equations are related to the sinh-Gordon formulae \cite{Ed}, by
analytic continuation, and this continuation also gives a definite \(
M_{crit}(\lambda ) \). There are some unpublished results on the
sine-Gordon case \cite{Zamm}, which also seem to give a similar
correspondence. At the moment, however, the relation between the
results in \cite{Ed} and \cite{Zamm} is not entirely clear to us, and
needs further clarification.

\section{Conclusions}

We determined the boundary spectrum for sine-Gordon model with the
most general integrable boundary condition, using the bootstrap
principle. We found that for the generic case the spectrum is
essentially identical to the one corresponding to Dirichlet boundary
conditions derived in \cite{patr}.

For all the poles of the breather and soliton reflection factors we
gave either an explanation as a boundary excited state or as a
boundary Coleman-Thun type diagram. It must be noted, however, that as
there exists no analogue of the Cutkosky rules for field theories with
boundaries, the boundary Coleman-Thun mechanism described in
\cite{bCT} at present is at most a reasonable guess which does seem to
work systematically in all the cases considered so far. Finding a
justification for the boundary rules in terms of a properly formulated
perturbative expansion together with a boundary extension of the
standard Landau rules for singularities of Feynman diagrams is a
challenge for the future.

We also note that for full consistency one must draw all the
Coleman-Thun diagrams and prove that their contributions add up to
explain the full residue of the given pole in the reflection
factors. This is very complicated in general and we have done it only
for some simple cases. For explicit examples see our previous paper
\cite{saj}, where we also noted the interesting fact that some poles
can only be explained by including contributions from both
Coleman-Thun type diagrams and boundary state creation processes,
which then fit consistently with other elements of the bootstrap.

Comparing the symmetries of the UV (Lagrangian) and IR (bound state
spectrum) description we gave arguments for a conjectured relationship
between their parameters.  This relationship can be confirmed by TCSA
or TBA analysis, and will be discussed in a separate publication
\cite{next}.

\subsubsection*{Acknowledgments}

G. T. thanks the Hungarian Ministry of Education for a Magyary Postdoctoral
Fellowship. This research was supported in part by the Hungarian Ministry
of Education under FKFP 0178/1999, 0043/2001 and by the Hungarian
National Science Fund (OTKA) T029802/99. 

\subsubsection*{Note added in proof}

After this paper was written we received information from P. Dorey
that P. Mattsson has already conjectured the same spectrum for
sine-Gordon theory. His results were unknown to us at that time, only
written down in a PhD thesis submitted to the University of Durham,
UK.  The full text of the thesis has recently been made available at
the Los Alamos preprint archive \cite{mthesis}.

\end{document}